\documentclass[12pt]{article}
\usepackage{graphicx}
\usepackage{dcolumn} 
\usepackage{bm}      
\usepackage{epsfig}
\usepackage{dcolumn}
\usepackage{amsmath}
\usepackage{cite}
\usepackage{changebar}

\def\be{\begin{equation}}
\def\ee{\end{equation}}
\begin{document}
\begin{center}
{\Large{\bf{Tsallis Statistics in High Energy Physics: Chemical and Thermal Freeze-Outs }}}

\vskip1cm

{\bf J.\ Cleymans$^1$,
M.\ W.\ Paradza$^{1,2}$}
\\[1cm]
1 UCT-CERN Research Centre and Physics Department, University of Cape Town, South Africa,\\
2 Centre for Postgraduate Studies, Cape Peninsula University of Technology, Bellville  7535, South Africa
\\

\bigskip

{\bf{Abstract:}
}
\end{center}
We present an overview of a proposal in relativistic proton-proton ($pp$) collisions emphasizing the thermal or kinetic freeze-out stage in the framework of the Tsallis distribution. In this paper we take into account the chemical potential present in the Tsallis distribution
by following a two step procedure. In the first step we used the
redundancy present in the variables such as the system temperature, $T$, volume, $V$, Tsallis exponent, $q$, chemical potential, $\mu$, and performed all fits by effectively
setting to zero the chemical potential.
In the second step the value $q$  is kept fixed at the value determined in the first step.
This way the complete  set of variables
$T, q, V$ and $\mu$ can be determined.
The final results show  a weak  energy dependence in $pp$ collisions at the centre-of-mass energy $\sqrt{s}= 6$~GeV to 13 TeV.
The chemical potential $\mu$ at kinetic freeze-out shows an 
increase with beam energy. This simplifies
the description of the thermal freeze-out stage in $pp$ collisions as the values of $T$ and of the freeze-out 
radius $R$ vary only mildly over a wide range of beam energies.

\newpage

~~~~~~~~~~
\vskip1cm

\section{Introduction}

It has been estimated~\cite{Adam:2016ddh}  that about 30,000 particles (pions, kaons, protons, antiprotons))
are produced in a central heavy ion collision at the Large Hadron Collider (LHC) at 5.02 TeV. Hence it is  natural to use concepts from statistical mechanics to analyze the produced particles.
This procedure has a long and proud history with contributions from three Nobel prize winners: E. Fermi~\cite{Fermi:1950jd, Fermi:1951zz},
W.~Heisenberg~\cite{Heisenberg:1952zz}
and L.D. Landau~\cite{Landau:1953gs}.
To quote Landau:
\begin{quote}
``{\emph{Fermi originated the ingenious idea of considering the collision process at very high energies by
the use of thermodynamic methods.}}''
\end{quote}

This turned out to be   useful also at much higher beam energies than those initially envisaged. The~main ingredient in  the hadron resonance gas model (referred to as  thermal model here)
is that all resonances  listed in the Review of Particle
Physics~\cite{Tanabashi:2018oca}  are in thermal and chemical equilibrium.
This  reduces the number of available parameters and just a few
thermodynamic variables characterize the~system.

The chemical freeze-out stage is well understood  and  is strongly supported by experimental
results~(see e.g.,~\cite{Andronic:2017pug} for a recent review) with  a strong connection to results obtained
using Lattice Quantum Chromodynamics (LQCD) as the chemical freeze-out temperature is consistent with
the phase transition temperature
calculated in LQCD.
Indeed, for the most central Pb-Pb collisions, the best description of the {ALICE}
data on
yields of particles in one unit of rapidity at mid-rapidity was obtained for a chemical freeze-out temperature
given by
$T_{ch} = 156.6\pm   1.7$ MeV \cite{Andronic:2017pug,Andronic:2018qqt}. Remarkably, this  value of  $T_{ch}$ is close to
the pseudo-critical temperature
$T_c = 156.5 \pm  1.5$  MeV obtained from first principles Lattice QCD  (LQCD) calculations \cite{Bazavov:2018mes}, albeit with
the possibility of a
broad transition region~\cite{Borsanyi:2020fev}.

For several  decades, a well-established procedure  using hydrodynamics~\cite{Sollfrank:1990qz} and variations
thereof has existed to
describe this stage.
In this paper we review another possibility to describe the thermal freeze-out stage which has shown considerable potential
especially to describe  the final state in
proton--proton  (pp) collisions. Most of these  approaches are based on variations  of a distribution
proposed by Tsallis about 40 years ago~\cite{Tsallis:1987eu} to describe entropy by introducing an additional parameter called $q$.
In the limit $q\rightarrow 1$ this reproduces the standard Boltzmann--Gibbs entropy. The~advantage is that thermodynamic
variables like temperature, energy density, pressure and particle density can still be used and thermodynamic consistency is maintained.

This paper is an extension of~\cite{Cleymans:2020nvs}. For completeness
and for the convenience of the reader we have included the tables presented there and considerably improved on them, the inclusion
of the NA61/SHINE~\cite{Abgrall:2013qoa} is new and contributes very much to the understanding of the
energy dependence of the parameters, also all figures are new.
\section{Thermal Freeze-Out}

We will focus here on
one particular form of the Tsallis distribution, satisfying thermodynamic consistency relations~\cite{Cleymans:2011in,Cleymans:2012ya}
and given by:
\begin{equation} \label{Yield}
E\frac{d^3N}{d^3p} = gV E\frac{1}{(2 \pi)^3} \left[  1 + ( q - 1) \frac{ E - \mu}{T}\right]^ {-\frac{q}{q -1}},
\end{equation}
where $V$ is the volume, $q$ is the Tsallis parameter, $T$ is the corresponding temperature, $E$ is the
energy of the particle, $p$ is the momentum, $g$ is the degeneracy factor and $\mu$ is the chemical potential.
In terms of variables  commonly used in high-energy physics,   rapidity $y$, transverse mass $ m_T = \sqrt{p_T^2 + m^2}$:
\begin{equation} \label{YieldNonZeroMu}
\frac{d^2N}{dp_Tdy}  = gV \frac{p_Tm_T\cosh y }{(2 \pi)^2} \left[  1 + ( q - 1) \frac{m_T \,\cosh\, y - \mu}{T}\right]^ {-\frac{q}{q -1}}.
\end{equation}

In the limit where the parameter $q$ tends to unity one recovers the well-known Boltzmann--Gibbs~distribution (with $p_T$ being the particle transverse momentum):
\begin{equation} \label{BG}
\lim_{q\rightarrow 1}\frac{d^2N}{dp_Tdy}  = gV \frac{p_Tm_T\cosh y }{(2 \pi)^2} \exp\left( - \frac{m_T \,\cosh\, y - \mu}{T}\right).
\end{equation}

The main advantage of Equation~(\ref{YieldNonZeroMu}) over Equation~(\ref{BG}) is that it has a polynomial decrease with increasing $p_T$
which is what is observed experimentally.

It was recognized early on~\cite{Cleymans:2013rfq} that there is a redundancy in the number of parameters in this distribution,
namely the four parameters $T,V,q$ and $\mu$ in Equation~$(\ref{YieldNonZeroMu})$ can be replaced by just three parameters $T_0,V_0,q$ with the
help of the following transformation:
\begin{eqnarray}
T_0 &=& T \left[1-(q-1) \frac{\mu}{T} \right], \qquad  \mu\leq\frac{T}{q-1},
\label{T0}  \\
V_0 &=& V  \left[1-(q-1) \frac{\mu}{T} \right]^{\frac{q}{1-q}},
\label{V0}
\end{eqnarray}
leading to a
transverse momentum distribution which can thus be written equivalently as
\begin{equation} \label{YieldIndexZero}
\frac{d^2N}{dp_Tdy}  = gV_0 \frac{p_Tm_T\cosh y}{(2 \pi)^2} \left[  1 + ( q - 1) \frac{m_T \,\cosh\, y}{T_0}\right]^ {-\frac{q}{q -1}},
\end{equation}
where the chemical potential does not appear explicitly.

Corresponding to the volumes $V$ and $V_0$ defined in Equations~(\ref{Yield}) and (\ref{V0}) we also introduce the corresponding
radii $R$ and $R_0$
\begin{eqnarray}
V &=& \frac{4\pi}{3}R^3  ,
\label{R0}\\
V_0 &=& \frac{4\pi}{3}R_0^3 \label{R} .
\end{eqnarray}

It is to be noted that most previous analyses have confused the two Equations~(\ref{YieldNonZeroMu})
and (\ref{YieldIndexZero}) and
reached
conclusions that are incorrect,  namely  that at LHC energies, different hadrons,
$\pi, K, p, ...$ cannot be described by the same values of $T$ and $V$. As we will show this is based on using $T_0$ and $V_0$
and not $T$ and $V$. Many authors have followed this conclusion because at LHC energies equal numbers of
particles and antiparticles are being produced and, furthermore, at chemical equilibrium, one has indeed $\mu = 0$ MeV
for all quantum numbers. However the equality
of particle and antiparticle yields, at thermal freeze-out,  only implies that e.g., $\pi^+$ and $\pi^-$ have the same chemical
potential but they are  not necessarily zero.
We emphasize that Equations~(\ref{YieldNonZeroMu}) and~(\ref{YieldIndexZero}) carry a different meaning, notice the
difference in parameters: $T_0$ is not equal to $T$ and neither is $V$ equal to $V_0$. Notice also that we do not have $\mu$ in
Equation~(\ref{YieldIndexZero}).

It is the purpose of the present paper to resolve this issue.
The procedure we choose is the~following:
\begin{enumerate}
\item Use Equation~(\ref{YieldIndexZero}) to fit the transverse momentum distributions. This determines the three parameters $T_0$, $q$ and $V_0$.
\item Fix the parameter $q$ thus obtained.
\item Perform a new fit to the transverse momentum distributions using Equation~(\ref{YieldNonZeroMu}) keeping $q$ as determined in the previous step.
This determines the parameters $T$ and $V$ and the chemical potential $\mu$.
\item Check the consistency with Equations~(\ref{T0}) and (\ref{V0}).
\end{enumerate}

Each step in the fitting procedure thus  involves only three parameters to describe the transverse momentum distributions.
This procedure was presented in~\cite{Cleymans:2020nvs} and the present paper is an extension with more details in this
paper,
some of the entries in Table 2 have been corrected.

We emphasize that the chemical potentials at kinetic freeze-out (described here with a Tsallis distribution), are not related to those
at chemical freeze-out.
At chemical freeze-out, where thermal and chemical equilibrium have been well established
the chemical potentials are zero.
At kinetic freeze-out however, there is no chemical equilibrium and the observed particle-antiparticle symmetry only
implies that the chemical potentials for particles must be equal to those for antiparticles.
However, due to the absence of chemical  equilibrium they do not have to be zero.
The only constraint is that they should be equal for particles and antiparticles.

We remind the reader here of
the advantage of using the above distribution as they follow a  consistent set of thermodynamic relations
(see e.g.,~\cite{Cleymans:2013rfq}).
From this, it is thus clear that the parameter $T$ can indeed be  considered as a temperature in the thermodynamic
sense since the relation below holds
\begin{equation}
T = \left.\frac{\partial E}{\partial S}\right|_{V,N},
\end{equation}
where the entropy $S$ is the Tsallis entropy.

In the next section we  include the chemical potential parameter in
the  Tsallis fits to the transverse momentum spectra. Previously, it was first
noted by~\cite{Cleymans:2013rfq}  that the variables $T,V,q$ and $\mu$ in the Tsallis distribution
function Equation~$(\ref{Yield})$ have a redundancy for $\mu \neq 0$ MeV and recently~\cite{biro2020tsallis} considered the mass of a particle in place of chemical potential. This necessitates work on determining the chemical potential from the transverse momentum spectra.

\section{Comparison of Fit Results}
As mentioned in the introduction, we reproduce here for completeness
the values extracted  from the results published by
the ALICE Collaboration~\cite{ALICE2,ALICE_2760,ALICE_5020,ALICE_7000}.
The data at the centre-of-mass energy  $\sqrt{s} = 0.9$  TeV had
the smallest range in $p_T$  (for all the ALICE Collaboration results considered here), of~about an
order of magnitude less than the experimental data  at  $\sqrt{s}  =  2.76$ GeV and  $\sqrt{s}  = 7$ TeV with ALICE.

In general the data were  described  very well; the figures showing the actual fits results are not
included in this paper since they form part of previous publications.
The least squares method was performed by the {Minuit} package~\cite{James:1975dr} as part of the fitting procedure in the code.
There was no manual selection in the choice of parameters, all parameters were initialized at the beginning and the code
returned the best fit parameter values. We did not particularly fix the value of $T$ and tried to obtain the other parameters.
In particular, the value of $\mu$ did not affect $V$.

We did not observe any trend which suggested a deterioration of the fits with the centre-of-mass energy.
In Tables \ref{ALICE_T0} and \ref{tab:results_fixedq}; we give the $\chi^2$  values.
Comparing the values of $\chi^2$  from 2.76 to 7.0 TeV, in~Tables \ref{ALICE_T0} and \ref{tab:results_fixedq}, there was no clear trend with increasing energy.

\begin{table}[ht]
\centering
\caption{{Fit results} at $\sqrt{s}$ = 0.9~\cite{ALICE2}, 2.76~\cite{ALICE_2760}, 5.02~\cite{ALICE_5020} and 7 TeV~\cite{ALICE_5020,ALICE_7000}, using data from the {ALICE}
Collaboration using Equations~(\ref{YieldIndexZero}) and~(\ref{R0}).}
\centering
\scalebox{.95}{\begin{tabular}{cccccc}\hline
\boldmath{$\sqrt{s} $ \textbf{(TeV)}}&\textbf{Particle}     & \boldmath{$R_0$} \textbf{(fm) }            & \boldmath{$q$ }                  & \boldmath{$T_0$} \textbf{(GeV) }     & \boldmath{$\chi^2$}\textbf{/NDF}          \\
\hline
0.9 &$\pi^+$ & 4.83 $\pm$ 0.14  & 1.148 $\pm$ 0.005 & 0.070 $\pm$ 0.002 & 22.73/30  \\
&$\pi^-$   & 4.74 $\pm$ 0.13     & 1.145 $\pm$ 0.005     & 0.072 $\pm$ 0.002    & 15.83/30  \\
&$K^+$     & 4.52 $\pm$ 1.30    & 1.175 $\pm$ 0.017     & 0.057 $\pm$ 0.013    & 13.02/24    \\
&$K^-$     & 3.96 $\pm$ 0.96    & 1.161 $\pm$ 0.016     & 0.064 $\pm$ 0.013    & 6.21/24   \\
&$p$       & 42.7 $\pm$ 19.8	 &  1.158 $\pm$ 0.006   &  0.020 $\pm$ 0.004  &  14.29/21 \\
&$\bar{p}$ & 7.44 $\pm$ 3.95   & 1.132 $\pm$ 0.014    & 0.052 $\pm$ 0.016     & 13.82/21 \\
\hline
2.76 &$\pi^+ + \pi^-$ & 4.80 $\pm$ 0.10 & 1.149 $\pm$ 0.002 & 0.077 $\pm$ 0.001    & 20.64/60  \\
&$K^+ + K^-$     & 2.51 $\pm$ 0.13  & 1.144 $\pm$ 0.002     & 0.096 $\pm$ 0.004  & 2.46/55    \\
&$p + \bar{p}$ & 4.01 $\pm$ 0.62   & 1.121 $\pm$ 0.005    & 0.086 $\pm$ 0.008     & 3.51/46 \\
\hline
5.02 &$\pi^+ + \pi^-$ & 5.02 $\pm$ 0.11 & 1.155 $\pm$ 0.002 & 0.076 $\pm$ 0.002    & 20.13/55  \\
&$K^+ + K^-$     & 2.44 $\pm$ 0.17  & 1.15 $\pm$ 0.005     & 0.099 $\pm$ 0.006  & 1.52/48    \\
&$p + \bar{p}$ & 3.60 $\pm$ 0.55   & 1.126 $\pm$ 0.005    & 0.091 $\pm$ 0.009     & 2.56/46 \\
\hline
7.0 &$\pi^+ + \pi^-$  & 5.66 $\pm$ 0.17 & 1.179 $\pm$ 0.003 & 0.066 $\pm$ 0.002   & 14.14/38  \\
&$K^+ + K^-$     & 2.51 $\pm$ 0.15  & 1.158 $\pm$ 0.005 & 0.097 $\pm$ 0.005    & 3.11/45    \\
&$p + \bar{p}$ & 3.07 $\pm$ 0.41   & 1.124 $\pm$ 0.005  & 0.101 $\pm$ 0.008   & 6.03/43 \\
\hline
\end{tabular}}
\label{ALICE_T0}
\end{table}\unskip

\begin{table}[ht]
\caption{ Fit results at $\sqrt{s}$ = 0.9~\cite{ALICE2},  2.76~\cite{ALICE_2760},
5.02~\cite{ALICE_5020} and 7 TeV~\cite{ALICE_5020,ALICE_7000}, using data from the
ALICE Collaboration with $q$ from Table~\ref{ALICE_T0} following Equations~(\ref{YieldNonZeroMu}) and~(\ref{R}).} 
\centering

\scalebox{.95}{\begin{tabular}{cccccc}\hline
\boldmath{$\sqrt{s}$ \textbf{(TeV)}}&\textbf{Particle}     & \boldmath{$R$} \textbf{(fm)}             & \boldmath{$\mu$} \textbf{(GeV)}                  & \boldmath{$T$} \textbf{(GeV)}      & \boldmath{$\chi^2$}\textbf{/NDF}          \\
\hline
0.9 &$\pi^+$    & 3.64 $\pm$ 0.21  & 0.055 $\pm$ 0.012   & 0.079 $\pm$ 0.002  & 3.66/30  \\
&$\pi^-$        & 3.53 $\pm$ 0.21  & 0.059 $\pm$ 0.012   & 0.080 $\pm$ 0.002  & 2.18/30  \\
&$K^+$          & 3.76 $\pm$ 0.33  & 0.029 $\pm$ 0.017   & 0.062 $\pm$ 0.003  & 5.31/24    \\
&$K^-$          & 3.89 $\pm$ 0.35  & 0.003 $\pm$ 0.018   & 0.065 $\pm$ 0.003  & 3.38/24   \\
&$p$            & 3.34 $\pm$ 0.27  & 0.233 $\pm$ 0.020   & 0.057 $\pm$ 0.007  & 7.44/21  \\
&$\bar{p}$      & 3.93 $\pm$ 0.33  & 0.097 $\pm$ 0.024   & 0.065 $\pm$ 0.002  & 7.69/21   \\
\hline
2.76 &$\pi^+ + \pi^-$  & 4.32 $\pm$ 2.68   &  0.022 $\pm$ 0.130  & 0.080  $\pm$ 0.019   &  20.48/60   \\
&$K^+ + K^-$           & 4.75 $\pm$ 0.03   & $-$0.140 $\pm$ 0.008  & 0.075  $\pm$ 0.004   & 2.48/55   \\
&$p + \bar{p}$         & 4.47 $\pm$ 5.50   & $-$0.071 $\pm$ 0.253  & 0.077  $\pm$ 0.030     & 3.52/46 \\
\hline
5.02 & $\pi^+ + \pi^-$ & 4.19 $\pm$ 2.64   & 0.038 $\pm$ 0.134  & 0.082 $\pm$ 0.021   & 20.14/55   \\
&$K^+ + K^-$           & 4.49 $\pm$ 0.03   & $-$0.142 $\pm$ 0.009  & 0.078 $\pm$ 0.0005  & 1.52/48   \\
&$p + \bar{p}$         & 4.00 $\pm$ 4.48   & $-$0.075 $\pm$ 0.243  & 0.081 $\pm$ 0.031  & 2.56/46   \\
\hline
7.0  &$\pi^+ + \pi^-$  & 3.67 $\pm$ 0.02   & 0.081 $\pm$ 0.141  & 0.081  $\pm$ 0.003   &  14.15/38   \\
&$K^+ + K^-$           & 3.80 $\pm$ 0.22   & $-$0.098$\pm$ 0.014  & 0.082  $\pm$ 0.002    & 3.13/55   \\
&$p + \bar{p}$         & 4.07 $\pm$ 0.27   & $-$0.127$\pm$ 0.018  & 0.085  $\pm$ 0.002    & 6.03/43 \\
\hline
\end{tabular}}

\label{tab:results_fixedq}
\end{table}

The fits to the transverse momentum distributions were then repeated using Equation~(\ref{YieldNonZeroMu}) but this time keeping the parameter $q$ fixed
to the value
determined in the previous section and listed in Table~\ref{ALICE_T0}. The~results are listed
in Table~\ref{tab:results_fixedq}, where we present the fit results for non-zero chemical potential
for $pp$ collisions at four different beam energies by the
ALICE Collaboration.

In the first case, we set the chemical potential as the mass of the respective particle and compare our results to~\cite{biro2020tsallis} for $pp$ collisions at $0.9$ TeV with the {CMS}
Collaboration, secondly we set the chemical potential as a free parameter to fit the data and analysis of the fit results and lastly, we calculated the chemical potential directly from Equation~$(\ref{poland_mu})$.

In Table~\ref{CMS_all_nonzero_mu} we present the extracted {values} of $T,\,\, q\, , R$ and $\mu$ at four different 
energies with the CMS Collaboration.

\begin{table}[ht]
\caption{\label{CMS_all_nonzero_mu}The extracted values of $T,\,\, q\, , R, \, \mu\,  \mathrm{and}\,\, \chi^2/NDF $ parameters, using the data published \mbox{in~\cite{Chatrchyan:2011av, Sirunyan:2017zmn,Chatrchyan:2012qb}} for $pp$ collisions with the CMS experiment.}
\centering
\begin{tabular}{llcccll}
\hline
\boldmath{$\sqrt{s}$} \textbf{(TeV)}         &   \textbf{Particle}	 & \boldmath{$T$}  \textbf{(MeV)}          	   & \boldmath{$ q$}                 &  \boldmath{$R$}  \textbf{(fm)}         & \boldmath{$\mu$} \textbf{(MeV)}            &  \boldmath{$\chi^2/NDF$}\\
\hline
$0.9$ \cite{Chatrchyan:2011av} & $\pi^+ $  & $ 77 \pm 1 $  & $1.164 \pm 0.004 $    & $ 0.070 \pm 0.102$  & $ 66 \pm 4$&  $ 8.111/18$\\
&$K^+ $      & $ 74\pm 1 $  & $1.158 \pm 0.008 $    & $ 3.724 \pm 0.126$  & $ -25 \pm 9 $&  $ 2.123/13$\\
& $p^+ $        & $ 71\pm 1 $  & $1.139\pm  0.003 $    & $3.536 \pm 0.105$   & $94 \pm 9$ &  $9.596/23$\\
\hline
$2.76$ \cite{Chatrchyan:2012qb} &$\pi^+ $  & $ 76\pm 1 $  & $1.189 \pm 0.005 $    & $ 3.906 \pm 0.100$ & $ 80 \pm 5$  &  $ 5.711/18$\\
& $K^+ $         & $ 78\pm 1 $  & $1.162 \pm 0.008 $    & $ 3.883 \pm 0.019$ & $ -5 \pm 1$ &  $ 2.447/13$\\
& $p^+ $         & $ 67\pm 1 $  & $1.166 \pm 0.004 $    & $ 3.508 \pm 0.099$ & $ 107 \pm 9$ &  $27.43/23$\\
\hline
$7.0$ \cite{Chatrchyan:2012qb} & $\pi^+ $ & $ 77\pm 1 $  & $1.203 \pm 0.005 $    & $ 3.994 \pm 0.105$ & $ 89 \pm 1$  &  $ 14.29/18$\\
& $K^+ $         & $87 \pm 1 $   & $1.152\pm 0.009 $    & $ 3.900 \pm 0.135$ & $ -96 \pm  11$ &  $ 2.074/13$\\
& $p^+ $        & $ 67 \pm 1 $  & $1.184 \pm 0.004 $    & $ 3.509 \pm 0.099$  & $ 84 \pm 9$   &  $12.22/23$\\
\hline
$13.0$ \cite{Sirunyan:2017zmn} & $\pi^+ $      & $ 76 \pm 2 $      & $1.215 \pm 0.008 $  & $ 3.932 \pm 0.157$  & $ 88 \pm 3$   &  $ 3.546/18$\\
& $K^+ $        & $ 88 \pm 3$      & $1.142 \pm 0.0150 $  & $ 4.044 \pm 0.27$   & $ -124 \pm  22 $   &  $ 1.828/13$\\
& $p^+ $        & $ 59 \pm 1$     & $1.213 \pm 0.008$   & $ 3.135 \pm 0.130$ & $191 \pm 14$    &  $8.892/22$\\
\hline
\end{tabular}
\end{table}

\begin{figure}[ht]
\centering
\includegraphics[height = 11cm, width = 12cm]{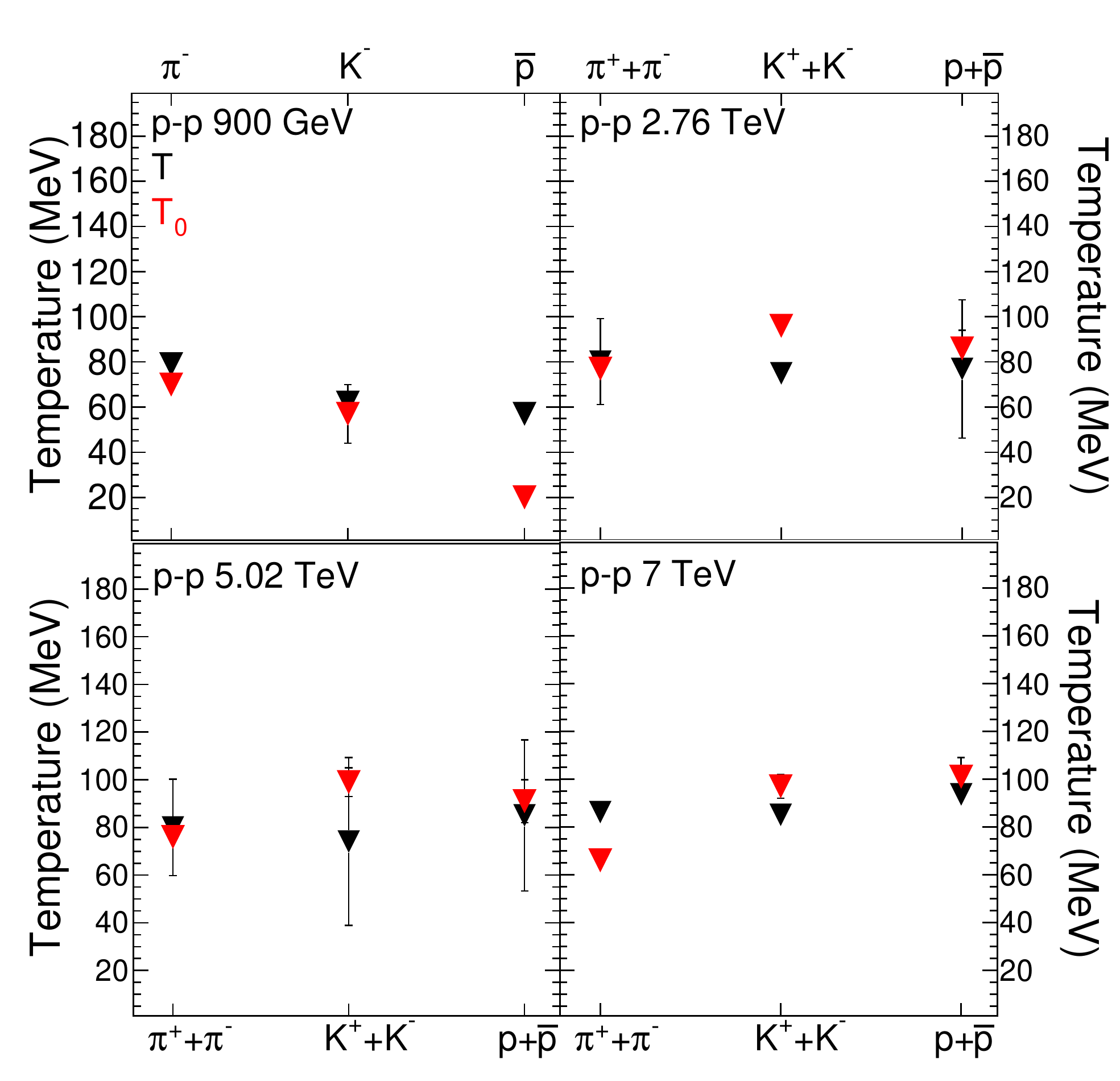} %
\caption{\label{T_and_T0_allenergies}A comparison of the values of temperatures $T$ and $T_0$ of different hadron species
for $pp$ collisions at $\sqrt{s}$ = 0.9~\cite{ALICE2}, 2.76~\cite{ALICE_2760}, 5.02~\cite{ALICE_5020} and $7$~\cite{ALICE_7000} TeV.
}
\end{figure}

In Figure~\ref{T_and_T0_allenergies} we  compare the  values for $T_0$ and $T$ at four different
beam energies. The~results obtained for $T$ were  more stable for different particle types
than the values obtained for $T_0$. We will come back to this with more detail later in this paper.

{An interesting proposal}  to determine the chemical potential was made in~\cite{Rybczynski:2014cha}, where 
the observation was made that the radius $R_0$ given  in Table~\ref{ALICE_T0} is larger than the one obtained from
a femtoscopy analysis~\cite{Aamodt:2011kd} by a factor $\kappa$ estimated to be about 3.5, i.e.,
\begin{equation}
R_{\rm femto} \approx \frac{1}{\kappa}R_0.
\end{equation}

Hence in~\cite{Rybczynski:2014cha} the   suggestion is made to identify the corresponding volume $V_{\rm femto}$ with the volume $V$ appearing in Equation~\eqref{Yield}.

Hence
\begin{equation}\label{PolandAssumption}
V_0 \approx V\cdot \kappa^3.
\end{equation}

Combining this with Equations~$(\ref{T0})$ and $(\ref{V0})$ this  leads to a chemical potential given by
\begin{equation} \label{poland_mu}
\mu = \frac{T_0}{q -1} \left( \kappa^{3(q -1)/q}   - 1  \right),
\end{equation}

Hence, using this proposal, a knowledge of $T_0$ would lead to a determination of $\mu$.

We compared the resulting values of the chemical potential $\mu$  using this proposal~\cite{Rybczynski:2014cha}
to the values using the procedure outlined above starting  Equation~\eqref{YieldNonZeroMu} and
concluded that the results are very different; hence, our results  do not
support this assumption
and thus the volume $V$ appearing in Equation~\eqref{YieldNonZeroMu} cannot be identified with the volume determined from
femtoscopy. The~volume $V$ must be considered to be  specific to the Tsallis distribution as is the case with all the other
variables used in this paper.

A clearer picture of the energy dependence emerges when including results from
the NA61/SHINE Collaboration~\cite{Abgrall:2013qoa} for
$\pi^-$'s.
The procedure outlined above was repeated in this case using the data published in~\cite{Abgrall:2013qoa}, first we
used Equation~\eqref{YieldIndexZero} and collect the results in Table~\ref{NA61_T0}. Next we fix the
values of $q$ obtained this way and repeat the fits using Equation~\eqref{Yield}; the
results are then collected in Table~\ref{NA61_T}.

\unskip
\begin{table}[ht]
\caption{\label{NA61_T0}The extracted values of $T_0,\,\, q\, , R_0\,\, \mathrm{and}\,\, \chi^2/NDF $  parameters, using
the data published in \cite{Abgrall:2013qoa} for $pp$ collisions with the  NA $61$ Collaboration.}
\centering
\begin{tabular}{cccccc} \hline
\boldmath{$\sqrt{s}$} \textbf{(GeV)} &   \textbf{Particle}	  & \boldmath{$T_0$}  \textbf{(MeV) }     	 & \boldmath{$ q$}                           &  \boldmath{$R_0$}  \textbf{(fm)}                 &  \boldmath{$\chi^2/NDF$}\\ \hline
$6.3$        &  $\pi^-$   & $ 98\pm 6 $        & $1.042 \pm 0.015 $   & $ 2.55 \pm 0.14$   &  $ 4.454/15$\\
$7.7$       &  $\pi^-$    & $ 95\pm 3 $  & $1.057 \pm 0.008 $   & $ 2.72  \pm 0.09$  &  $ 4.561/15$\\
$8.8$  		& $\pi^-$    & $ 96\pm 2 $  & $1.055 \pm 0.006 $   & $ 2.76 \pm 0.06$   &  $8.423/15$\\
$12.3$     	&  $\pi^-$   & $ 95\pm 2 $  & $1.064 \pm 0.006 $   & $ 2.90 \pm 0.06$   &  $ 6.775/15$\\
$17.3$   	&  $\pi^-$   & $ 93\pm 3 $  & $1.069 \pm 0.006 $   & $ 3.07  \pm 0.08$   &  $ 2.176/15$\\
\hline
\end{tabular}
\end{table}
\unskip
\begin{table}[ht]
\caption{\label{NA61_T}The extracted values of $T,\,\, \mu\, , R\,\, \mathrm{and}\,\, \chi^2/NDF $
parameters, using
the data published in \cite{Abgrall:2013qoa} for $pp$ collisions with the  NA $61$ Collaboration.}
\centering
\begin{tabular}{cccccc} \hline
\boldmath{$\sqrt{s}$} \textbf{(GeV)} &   \textbf{Particle}	  & \boldmath{$R$}  \textbf{(fm)}  & \boldmath{$ \mu$} \textbf{(GeV)}        &  \boldmath{$T$ } \textbf{(GeV)}  &  \boldmath{$\chi^2/NDF$}\\ \hline
$6.3$       &  $\pi^-$   & $ 2.451 \pm 0.399 $ & $ 0.011 \pm 0.046 $   & $ 0.098 \pm 0.003$   &  $ 4.454/15$\\
$7.3$       &  $\pi^-$   & $ 2.529 \pm 0.223 $ & $ 0.020 \pm 0.024 $   & $ 0.096 \pm 0.002$   &  $ 4.561/15$\\
$8.8$  	   &  $\pi^-$   & $ 2.548 \pm 0.016 $ & $ 0.022 \pm 0.002 $   & $ 0.097 \pm 0.001$   &  $ 8.423/15$\\
$12.3$       &  $\pi^-$   & $ 2.638 \pm 0.171 $ & $ 0.026 \pm 0.018 $   & $ 0.096 \pm 0.001$   &  $ 6.776/15$\\
$17.3$      &  $\pi^-$   & $ 2.785 \pm 0.216 $ & $ 0.025 \pm 0.021 $   & $ 0.095 \pm 0.002$   &  $ 2.179/15$\\
\hline
\end{tabular}
\end{table}

\begin{figure}[ht]
\centering
\includegraphics[height = 8.8cm, width = 10cm]{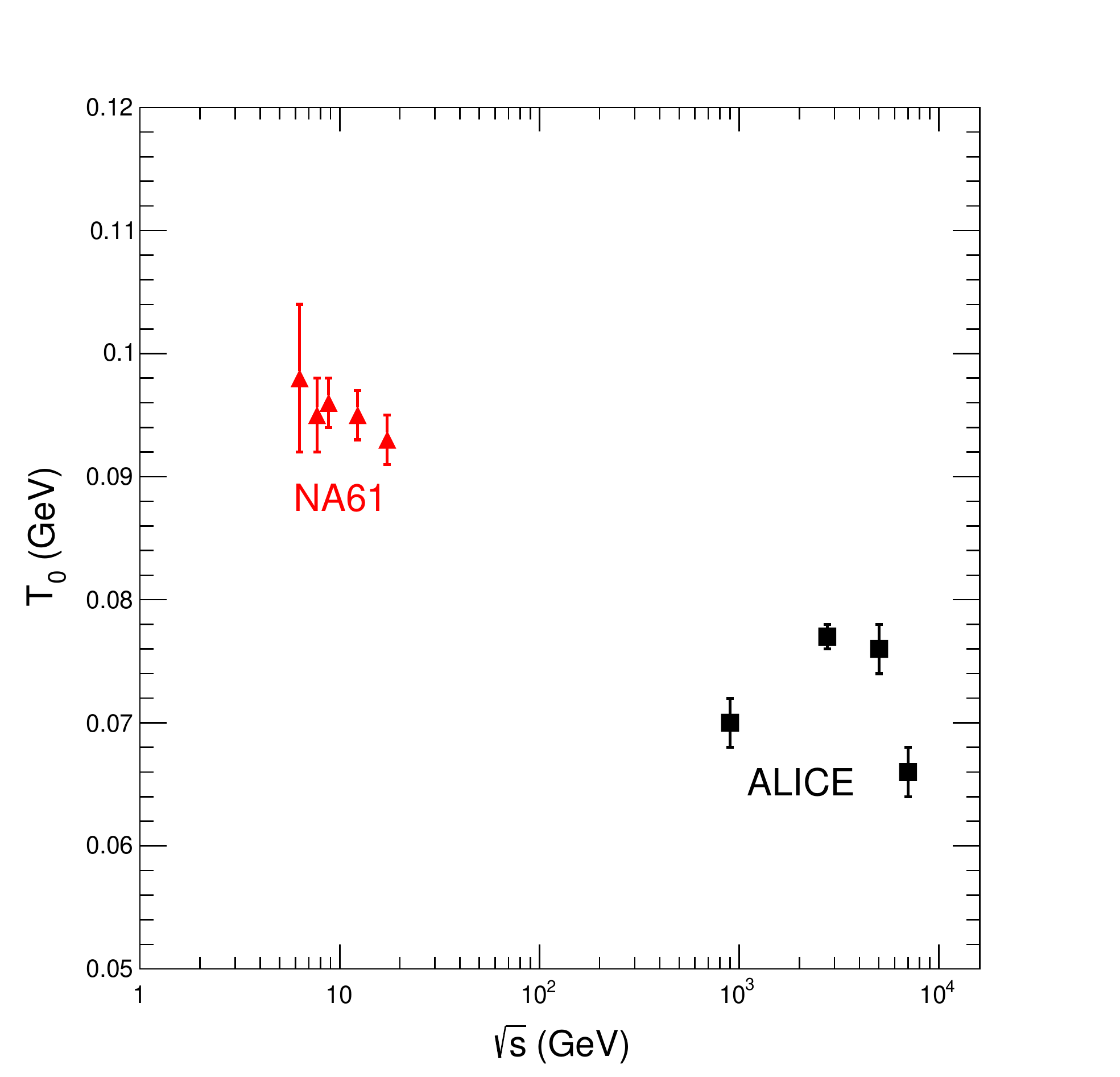} %
\caption{\label{NA61_ALICE_T0}The energy dependence of the temperature parameter $T_0$.
The triangular points are values of $T_0$ extracted from data in  $pp$ collisions
obtained by  the NA61/SHINE~\cite{Abgrall:2013qoa} 
Collaboration for pions (see~Table~\ref{NA61_T0}).
The  squares are the $T_0$ values in  Table~\ref{ALICE_T0}.
All points were obtained by fits using  Equation~(\ref{YieldIndexZero}).}
\end{figure}

The values for $T_0$ as a function of beam energy are shown in Figure~\ref{NA61_ALICE_T0}.
As one can see  a fairly strong energy dependence was present when comparing the two sets of data.

\begin{figure}[ht]
\centering
\includegraphics[height =8.8cm, width = 10cm]{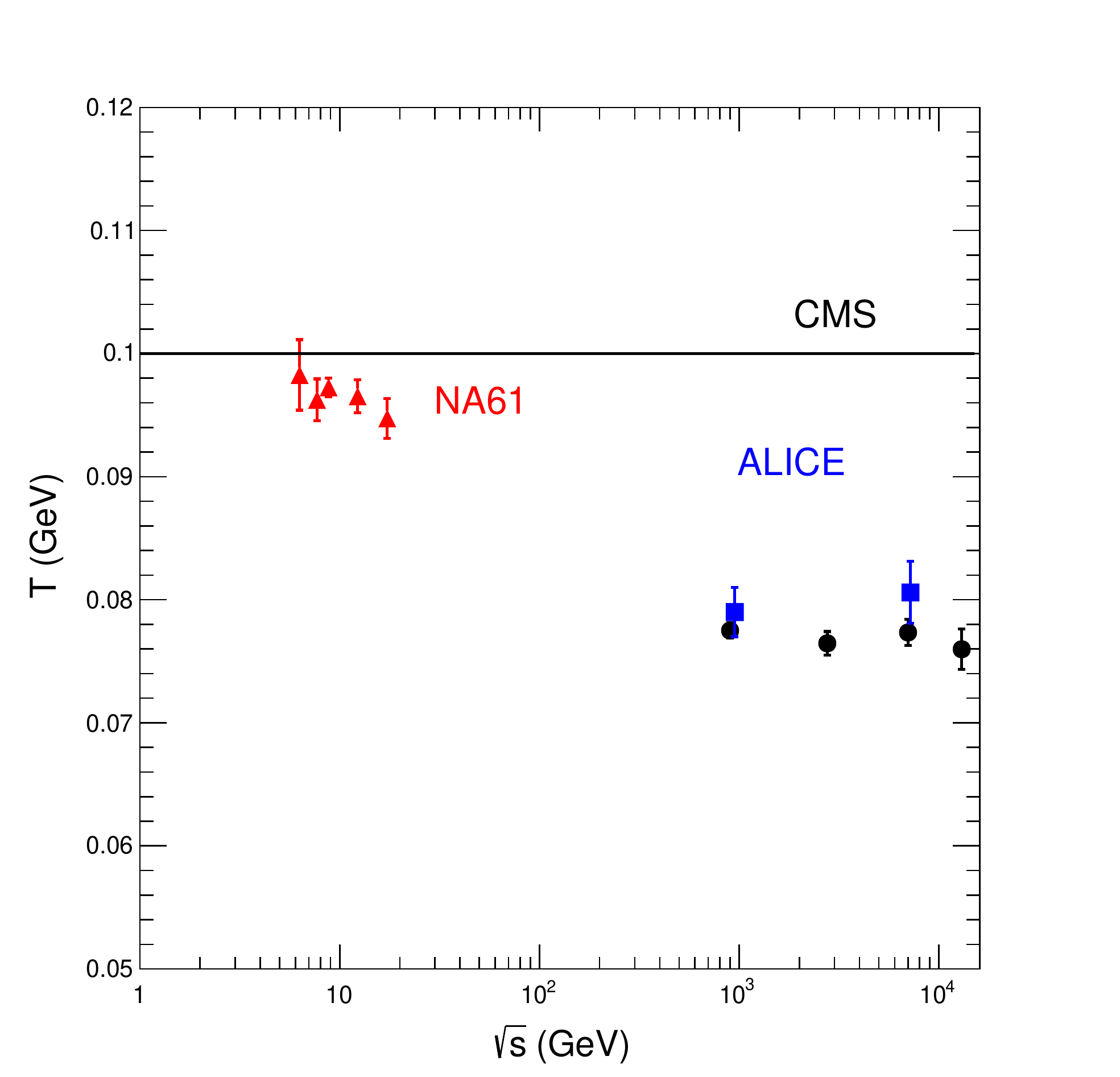} %
\caption{\label{NA61_ALICE_CMS_T} The energy dependence of the temperature $T$ for pions in $pp$
collisions. The triangular points are values of $T$ extracted from data in  $pp$ collisions
obtained by  the NA61/SHINE~\cite{Abgrall:2013qoa}
Collaboration for pions (see  Table~\ref{NA61_T}).
The squares are the $T$ values in Tables ~\ref{tab:results_fixedq} and \ref{CMS_all_nonzero_mu}.
All points were obtained by fits using  Equation~(\ref{YieldNonZeroMu}).
The straight line at $T$ =  0.1 GeV is there to guide the eye only.}
\end{figure}

However, this picture changes when plotting the temperature $T$ as a function of beam energy
as shown in Figure~\ref{NA61_ALICE_CMS_T}. The~energy dependence becomes weaker  and the  values of $T$ decrease
with increasing beam energy from about 10 GeV all the way up to 13,000 GeV. A similar decrease of the
kinetic freeze-out energy was also observed by the STAR collaboration~\cite{Adam:2019koz} at the Brookhaven
National Laboratory.

\begin{figure}[ht]
\centering
\includegraphics[height = 8.5cm, width = 10cm]{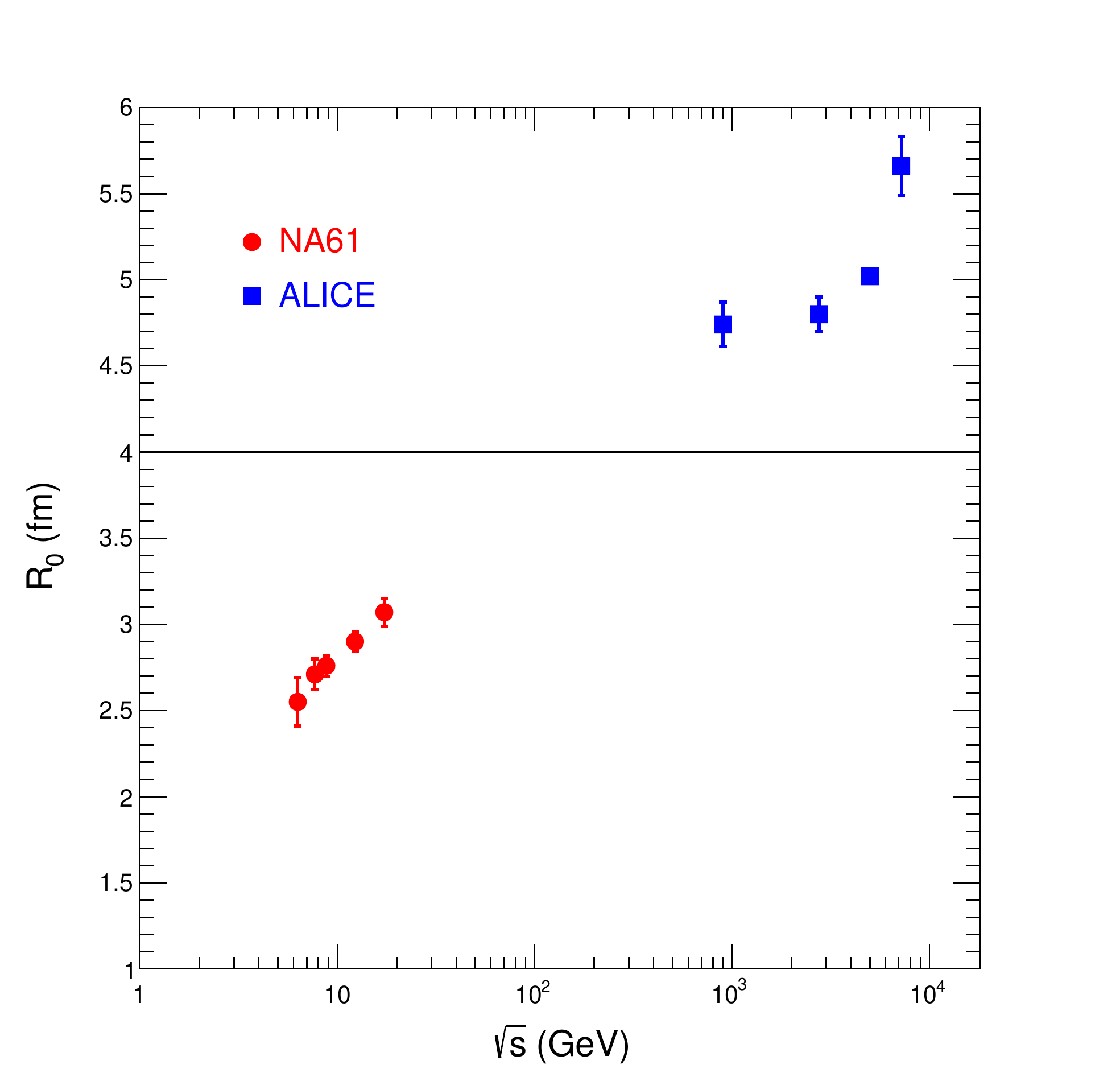} %
\caption{\label{NA61_ALICE_CMS_R0} The energy dependence of the freeze-out radius $R_0$ of pions in $pp$ collisions.
The round (red) points are
obtained from fits to the results of the NA61/SHINE Collaboration~\cite{Abgrall:2013qoa} (see Table \ref{NA61_T0}), the~square points are for the ALICE Collaboration data (see Table \ref{ALICE_T0}).
The straight line at $R_0$ = 4 fm is there to guide the eye only.}
\end{figure}

\unskip
Similarly when plotting the results obtained for the radius $R_0$ one sees a strong dependence on the beam  energy
as seen in Figure~\ref{NA61_ALICE_CMS_R0}.\vspace{-3pt}

\begin{figure}[ht]
\centering
\includegraphics[height = 8.5cm, width = 10cm]{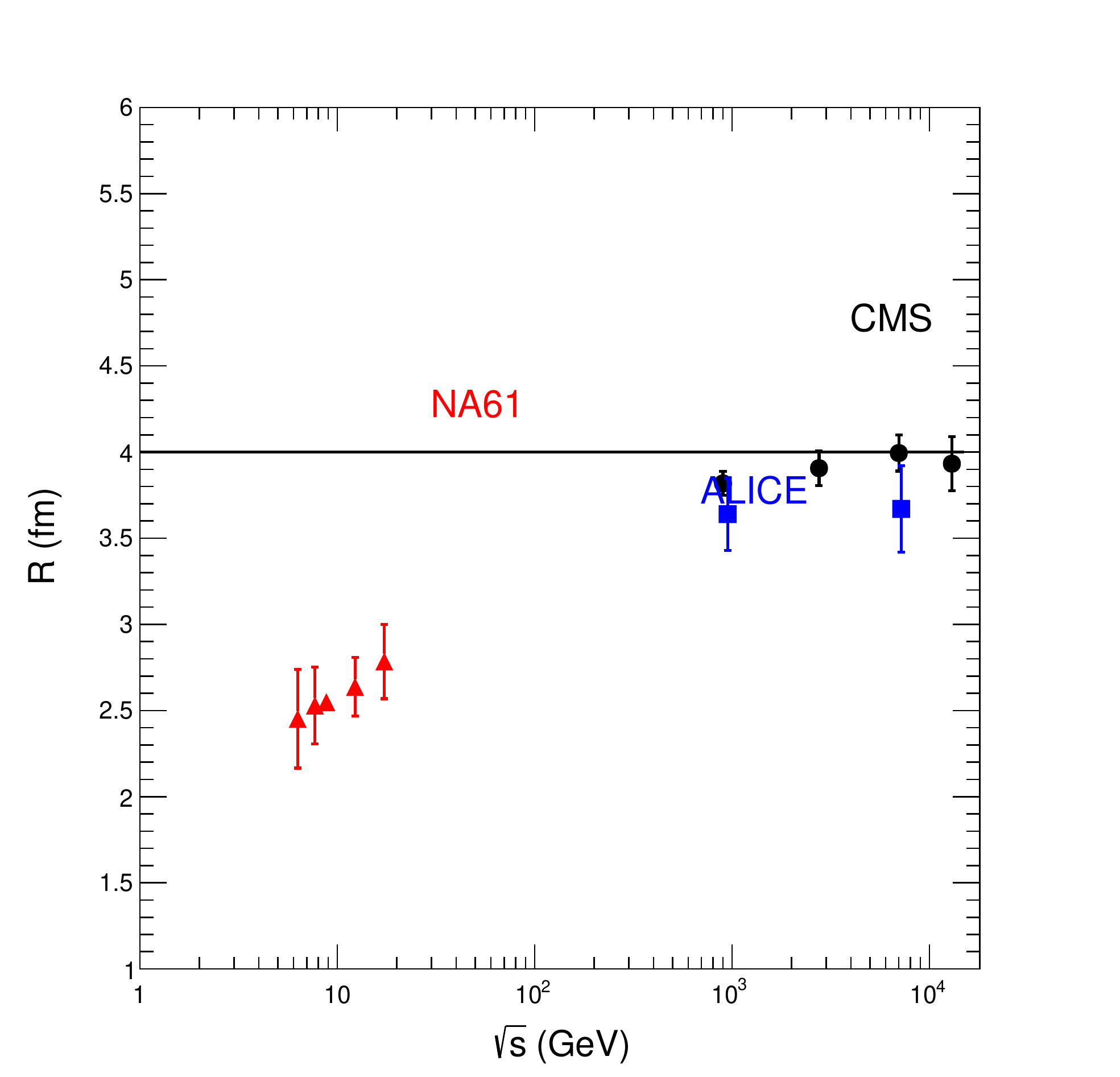} %
\caption{\label{NA61_ALICE_CMS_R} {The energy dependence of the} freeze-out radius $R$ of pions in $pp$ collisions.
The round (red) points are
obtained from fits to the results of the NA61/SHINE Collaboration~\cite{Abgrall:2013qoa} (see Table \ref{NA61_T}), the~square (blue) points are for the ALICE Collaboration data (see Table \ref{tab:results_fixedq}),
{while the round (black) points are for the CMS Collaboration data (see Table \ref{CMS_all_nonzero_mu}).
The straight line at $R$ = 4 fm is there to guide the eye only.}}
\end{figure}

However, similarly to the case with the temperatures $T$ and $T_0$, the energy dependence s weakened when plotting
the radius $R$ where only a very mild energy dependence could be noticed, see Figure \ref{NA61_ALICE_CMS_R}.\vspace{-3pt}

\begin{figure}[ht]
\centering
\includegraphics[height =9.5cm, width = 10cm]{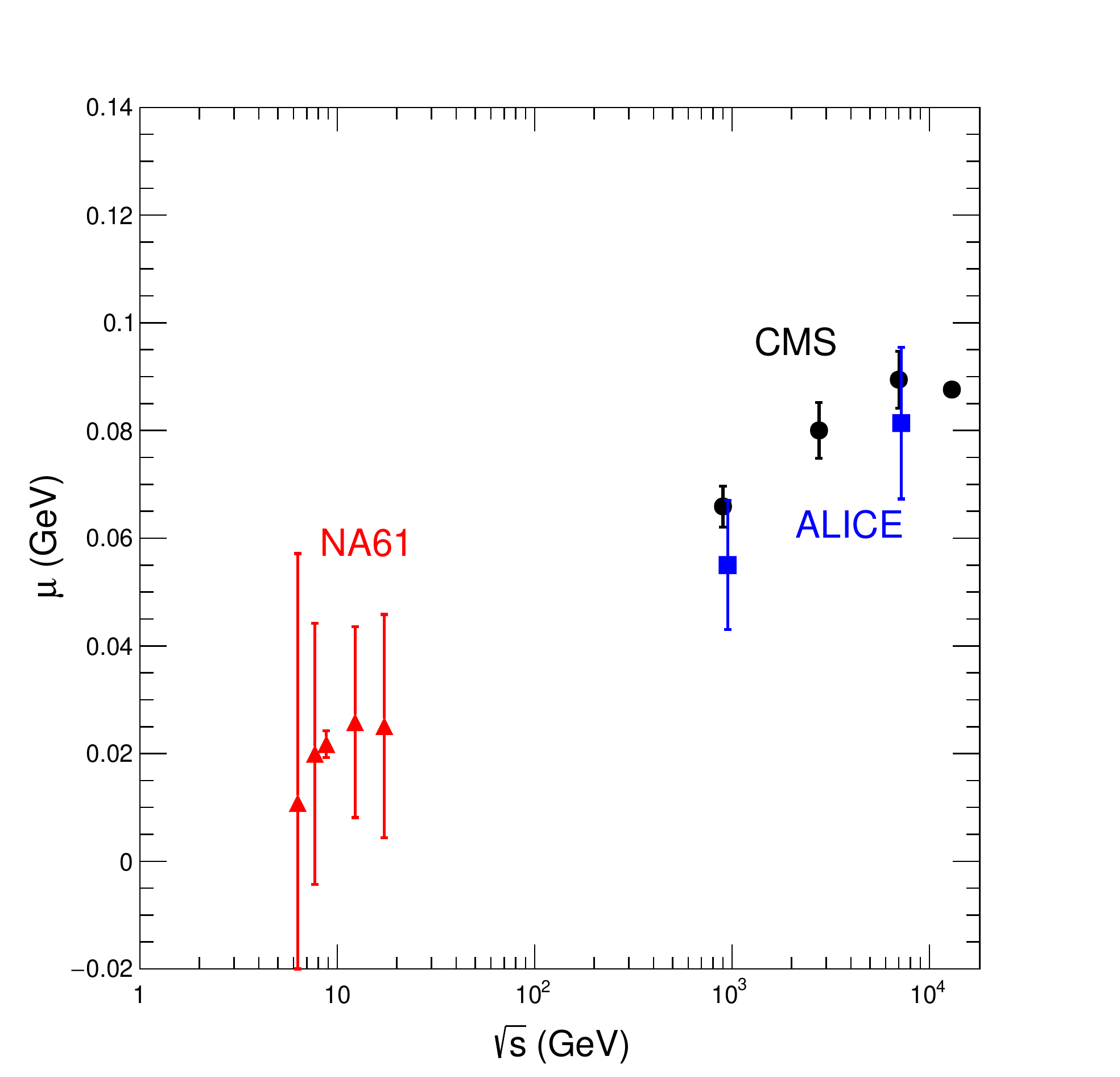} %
\caption{\label{NA61_ALICE_CMS_mu} The energy dependence of the freeze-out chemical potential $\mu$ for pions
in $pp$ collisions. The~round (red) points are
obtained from fits to the results of the NA61/SHINE Collaboration~\cite{Abgrall:2013qoa} (see Table \ref{NA61_T}), the square (blue) points
are for the ALICE Collaboration data (see Table \ref{tab:results_fixedq}),
while the round (black) points are for the CMS Collaboration data (see Table \ref{CMS_all_nonzero_mu}).
}
\end{figure}

Finally the parameter which was most influenced by deviations from chemical equilibrium, namely the chemical potential $\mu$
which is shown in Figure~\ref{NA61_ALICE_CMS_mu}. Here one sees a very clear increase with beam energy.

\section{Conclusions}
In this paper we have taken into account the chemical potential present in the Tsallis distribution Equation~(\ref{Yield})
by following a two step procedure. In the first step we used the redundancy present in the variables $T, V, q$ and $\mu$
expressed in Equations~\eqref{T0} and \eqref{V0} and performed all fits using Equation~\eqref{YieldIndexZero}, i.e.,~effectively
setting the chemical potential equal to zero.  The only variable which is common between Equations~\eqref{Yield} and
\eqref{YieldIndexZero} is the Tsallis parameter $q$; hence, in the second step of our procedure we
fixed the value of  $q$ and performed all fits using Equation~\eqref{Yield}. This way we finally obtained the set of variable
$T, V$ and $\mu$. The~results are shown in several figures. ~It is to be noted that $T$ and $R$ (as deduced
from the volume $V$) show a weak energy dependence in proton-proton ($pp$) collisions at the centre-of-mass energies
 from $\sqrt{s}$ = 20 GeV up to 7 and 13 TeV. This~is
not the case for the variables $T_0$ and $V_0$.  The chemical potential at kinetic freeze-out shows an
increase with beam energy as presented in Figure~\ref{NA61_ALICE_CMS_mu}. This simplifies
the resulting description of the thermal freeze-out stage in $pp$ collisions as the values of $T$ and $R$
vary mildly over a wide range of beam energies.

\vspace{6pt}

\end{document}